\documentclass{aa}
\usepackage[varg]{txfonts}
\usepackage[utf8]{inputenc}
\usepackage{comment}
\usepackage{amsmath}
\usepackage{graphicx}
\usepackage{gensymb}
\usepackage[colorlinks=true,linkcolor=blue,citecolor=blue]{hyperref}
\usepackage{xspace}
\newcommand{\srii}{\ion{Sr}{ii}\xspace}
\newcommand{\yii}{\ion{Y}{ii}\xspace}
\newcommand{\rprocess}{\textit{r}-process\xspace}

\usepackage{natbib}
\bibpunct{(}{)}{;}{a}{}{,} 

\begin{document}

\title{Measuring the Hubble constant with kilonovae using the expanding photosphere method} 
\titlerunning{Measuring the Hubble constant with kilonovae using the expanding photosphere method}
\authorrunning{A. Sneppen et al.}
\author{Albert Sneppen\inst{\ref{addr:DAWN},\ref{addr:jagtvej}} \and
Darach Watson\inst{\ref{addr:DAWN},\ref{addr:jagtvej}} \and 
Dovi Poznanski\inst{\ref{addr:Tel-aviv},\ref{addr:Caltech}} \and
Oliver Just\inst{\ref{addr:GSI},\ref{addr:RIKEN}} \and
Andreas Bauswein\inst{\ref{addr:GSI},\ref{addr:HFHF}} \and
Rados\l aw Wojtak \inst{\ref{addr:Dark}}
}

\institute{Cosmic Dawn Center (DAWN)\label{addr:DAWN}
\and
Niels Bohr Institute, University of Copenhagen, Blegdamsvej 17, K{\o}benhavn 2100, Denmark\label{addr:jagtvej} 
\and
School of Physics and Astronomy, Tel Aviv University, Tel Aviv, 69978, Israel\label{addr:Tel-aviv}
\and
Cahill Center for Astrophysics, California Institute of Technology, 1200 E. California Boulevard, Pasadena, CA 91125, USA \label{addr:Caltech}
\and
GSI Helmholtzzentrum f\"ur Schwerionenforschung, Planckstraße 1, D-64291 Darmstadt, Germany \label{addr:GSI}
\and
Astrophysical Big Bang Laboratory, RIKEN Cluster for Pioneering Research, 2-1 Hirosawa, Wako, Saitama 351-0198, Japan \label{addr:RIKEN}
\and
Helmholtz Forschungsakademie Hessen f{\"u}r FAIR, GSI Helmholtzzentrum für Schwerionenforschung, Planckstraße 1, 64291 Darmstadt, Germany \label{addr:HFHF}
\and
DARK, Niels Bohr Institute, University of Copenhagen, Jagtvej 128, 2200 Copenhagen, Denmark \label{addr:Dark}
}


\date{Received date /
Accepted date }

\abstract{
    While gravitational wave (GW) standard sirens from neutron star (NS) mergers have been proposed to offer good measurements of the Hubble constant, we show in this paper how a variation of the expanding photosphere method (EPM) or spectral-fitting expanding atmosphere method, applied to the kilonovae (KNe) associated with the mergers, can provide an independent distance measurement to individual mergers that is potentially accurate to within a few percent. There are four reasons why the  KN-EPM overcomes the major uncertainties commonly associated with this method in supernovae: 1) the early continuum is very well-reproduced by a blackbody spectrum, 2) the dilution effect from electron scattering opacity is likely negligible, 3) the explosion times are exactly known due to the GW detection, and 4) the ejecta geometry is, at least in some cases, highly spherical and can be constrained from line-shape analysis. 
    We provide an analysis of the early VLT/X-shooter spectra AT2017gfo showing how the luminosity distance can be determined, and find a luminosity distance of $D_L = 44.5\pm0.8$\,Mpc in agreement with, but more precise than, previous methods. We investigate the dominant systematic uncertainties, but our simple framework, which assumes a blackbody photosphere, does not account for the full time-dependent three-dimensional radiative transfer effects, so this distance should be treated as preliminary. The luminosity distance corresponds to an estimated Hubble constant of $H_0 = 67.0\pm 3.6$\,km\,s$^{-1}$\,Mpc$^{-1}$, where the dominant uncertainty is due to the modelling of the host peculiar velocity. We also estimate the expected constraints on $H_0$ from future KN-EPM-analysis with the upcoming O4 and O5 runs of the LIGO collaboration GW-detectors, where five to ten similar KNe would yield 1\% precision cosmological constraints. \newline
}

\maketitle
\section{Introduction}
A precise measurement of the Hubble constant, $H_0$, is critical to the study of cosmology. With the growing precision of cosmological studies has followed an increasing tension, currently significant at $5\sigma$, between measurements of the $H_0$ from type Ia 
supernovae (SNe) calibrated with Cepheids observed by the SH0ES collaboration \citep[for Supernova H0 for the Equation of State][]{Riess2021} and Planck observations of the cosmic microwave background radiation assuming a flat $\Lambda$CDM cosmological model \citep{Planck2018}. If these measurements are correct, this would suggest the standard cosmological paradigm, Lambda-cold dark matter ($\Lambda$CDM), cannot simultaneously fit observations at all redshifts, necessitating modifications to the cosmological model or to the physics of the early Universe, perhaps in the form of early dark energy \cite[e.g.][]{Poulin2019,Kamionkowski2022}. 

The $\Lambda$CDM paradigm may still be valid if unaccounted-for systematic effects have biased the inferred $H_0$ from either framework. For the local $H_0$ constraints this includes, but is not limited to, issues in modelling extinction in type Ia SNe \cite{Wojtak2022} or Cepheids \citep{Mortsell2022}, Cepheid metallicity correction \citep{Efstathiou2020}, host-galaxy properties \citep{Steinhardt2020}, and different populations of SNe\,Ia at low-$z$ and high-$z$ \citep{Jones2018,Rigault2020}. 
Given the many potential systematics, high-quality independent measurements of the Hubble constant on local cosmic distance scales can provide crucial evidence in the interpretation of the $H_0$ divergence.

Many independent probes have been developed, including time-delays of gravitationally lensed, variable sources \citep{Birrer2020,Wong2020}, megamaser distance measurements \citep{Pesce2020}, active galactic nucleus (AGN) dust reverberation \citep{Honig2014,Yoshii2014}, $\gamma$-ray attenuation by extragalactic background light \cite{Wojtak2019}, and the gravitational wave (GW) standard siren \citep{Schutz1986,Abbott2017a}. However, any conclusive interpretation remains elusive due to limited statistics and/or modelling uncertainties for all these methods. It has been argued that a rapid growth in GW sensitivity and an abundance of optical counterparts with redshifts could enable 1$\sigma$ constraints at the 2\% level using neutron star (NS) merger standard sirens within a few years \citep{Chen2018}. The tightest current $H_0$ constraint comes from combining very-long-baseline interferometry constraints with the GW standard siren, yielding fractional uncertainty of 7\% \citep{Hotokezaka2019}. It has also been argued that using the optical transient kilonova (KN) associated with NS merger events it might be possible to provide estimates of $H_0$ \citep{Doctor2020,Coughlin2020}, though constraints are comparatively poor and model-dependent compared to the standard siren. 

A P~Cygni feature associated with a transition of \srii was identified in the spectra of AT2017gfo, the kilonova associated with GW170817 \citep{Watson2019}. An additional P~Cygni feature from transitions in \yii was subsequently identified in later epochs \citep{Sneppen2023b}. These P~Cygni features have provided the first positive identification of freshly formed \rprocess elements, and allows the velocity and geometry of the kilonova to be probed. A recent analysis \citep{Sneppen2023} shows that the outer ejecta in AT2017gfo is spherically symmetric, which is not a generic outcome in existing theoretical models of merger ejecta \citep[e.g.][]{Kasen2015,Martin2015,Collins2023,Just2023}. Future studies will need to explore in detail the exact conditions under which the ejecta becomes highly spherical.
However, this sphericity, in conjunction with the clear blackbody continuum of early spectra does make AT2017gfo and similar kilonovae potentially excellent high-precision cosmological probes. 

We show in Sect.~\ref{sec:dl} how a variant of the expanding photosphere method \citep[EPM,][]{Kirshner1974} applied to KNe provides a novel and self-consistent estimator of cosmological distances with 2\% percent fractional uncertainty to AT2017gfo. In Sect.~\ref{sec:h0}, we present the corresponding $H_0$ constraints, which, as they are dominated by peculiar velocity uncertainties, remain broadly consistent with both late  and early Universe measurements. We discuss systematic effects in Sect.~\ref{sec:sys}. 
However, it is not our intent in this paper to make a definitive estimate of $H_0$ because we do not fully account for all spectral effects or for the major time-delay effects, which require more sophisticated and self-consistent 2D or 3D  radiative transfer modelling. Rather, we intend to show that a very accurate measurement of $H_0$ may be made using the kilonova AT2017gfo, and we show in Sect.~\ref{sec:future} how the highly constrained distance suggests that a handful of future kilonova detections may provide sub-percent precision constraints on $H_0$ entirely independent of the cosmic distance ladder.


\section{Methodology}\label{sec:methods}
The series of spectra of AT2017gfo taken with the medium-resolution ultraviolet (UV) to near-infrared (NIR) spectrograph, X-shooter, mounted at the Very Large Telescope at the European Southern Observatory, provides the most detailed information of any kilonova to date. Spectra were taken daily from 1.43 days after the event for 19 days, and show a temporal evolution of continuum, emission, and absorption features \citep{Smartt2017,Pian2017}. The spectra we use here are the same as those used in \citet{Sneppen2023} and follow the data reduction presented there and in \citet{Pian2017}. 
In the following section we deliberate on the spectral properties of both the blackbody continuum and the P~Cygni profiles of \srii, and then review the methodology and assumptions associated with the expanding photosphere method in Sect.~\ref{sec:epm}. 

\subsection{Blackbody continuum of AT2017gfo}\label{sec:bb}
Previous studies have established that the spectra from the earliest epochs of AT2017gfo are well-approximated by a blackbody spectrum \citep{Malesani2017,Waxman2018,Villar2017,Cowperthwaite2017,Nicholl2017,Arcavi2018,Drout2017,Shappee2017,Watson2019}. For a blackbody, the specific luminosity of the photosphere is set by the emitting area (which for a spherical expansion is set by the radius of the photosphere,  $4 \pi R_{ph}^2$), the Planck function $B(\lambda,T_{\rm eff})$ at a temperature $T_{\rm eff}$, and the relativistic Doppler and time-delay correction ($f(\beta)$) for a photosphere expanding with a velocity $\beta = v_{\rm ph}/c$: 
\begin{equation}
    L_{\lambda}^{BB} =  4 \pi R_{ph}^2 \pi  B(\lambda,T_{\rm eff}) f(\beta). 
    \label{eq:Lumino}
\end{equation}
Notably, for SNe the photospheric radius is above the thermalisation radius, due to the large contribution of electron-scattering opacity to the total opacity \citep{Eastman1996,Dessart2005,Dessart2015}. However, in KN ejecta the electron scattering opacity is small compared to the opacity of bound-bound transitions. This ratio is lower for two reasons. First, the opacity of bound-bound transitions is higher by several orders of magnitude due to the multiplicity of lines of \rprocess elements and the complexity of valence electron structure of elements within the \rprocess composition \citep{Kasen2013}, though this is somewhat lower for the lighter \rprocess elements. Secondly, the KN ejecta is composed of typically neutral singly- or doubly-ionised heavy elements, so the number of free electrons per unit mass is $O(10^{2})$ smaller than the typical value for ionised hydrogen. Thus, for the analysis of kilonovae we can assume the photospheric radius derived from the line is indeed at the thermalisation depth of the blackbody and no significant correction or `dilution factor' for electron opacity is required to determine the luminosity or distance.

\begin{figure}
    \includegraphics[width=\linewidth,viewport=15 15 485 560,clip=]{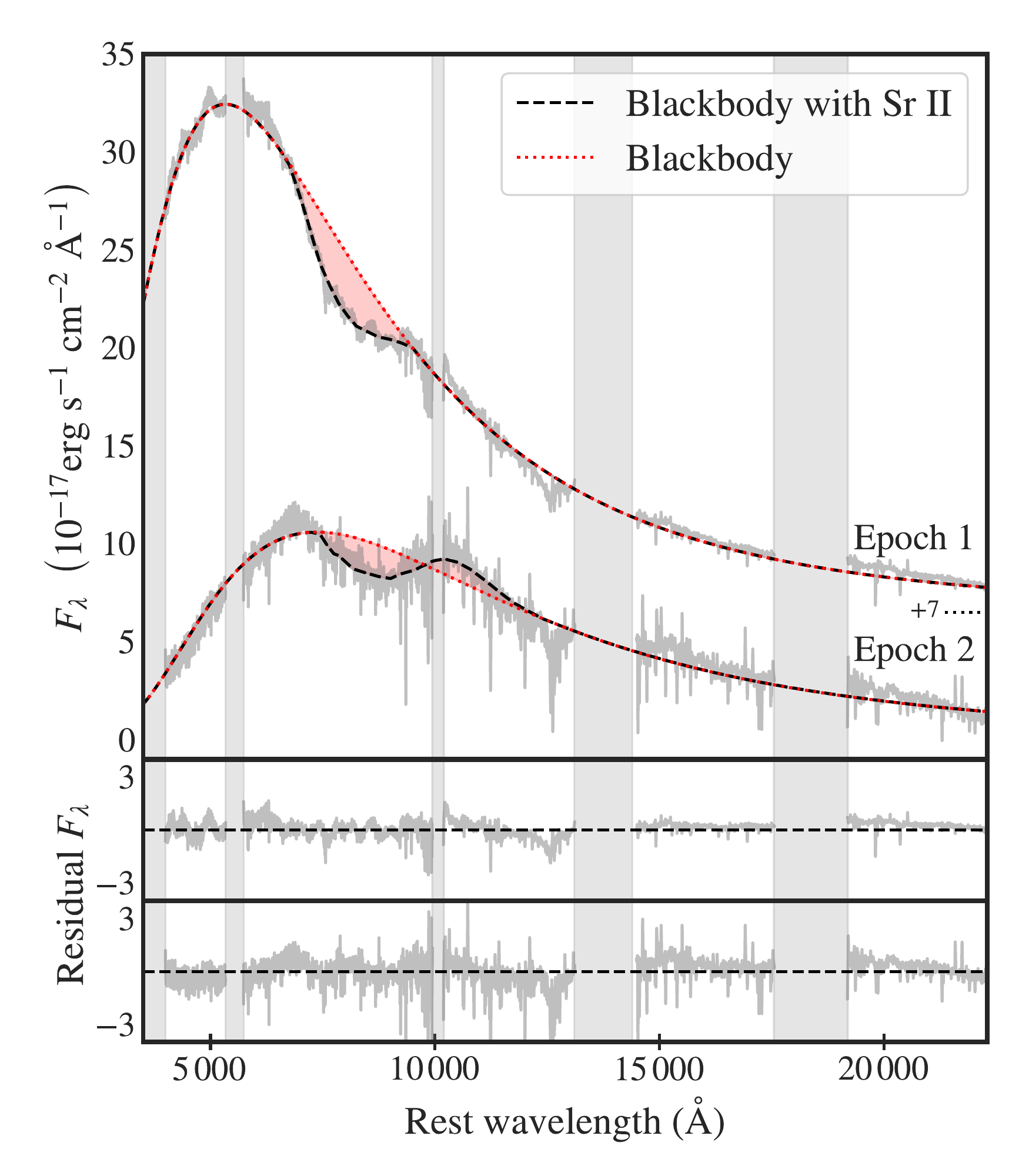}
    \caption{Spectrum and fit of the kilonova AT2017gfo spectrum for epochs 1 and 2 (days +1.43 and +2.42). The dotted line on the right side indicates the offset (\(7\times10^{-17}\)\,erg\,cm\(^{-2}\)\,s\(^{-1}\)\,\AA\(^{-1}\))
    added to the first epoch spectrum, the dark shaded bars indicate telluric regions, and the  light shaded bars indicate overlapping noisy regions at the edges of the UVB, VIS, and NIR arms.}
    \label{fig:X-shooter-spec-4}
\end{figure}

\subsection{P~Cygni modelling}\label{sec:Pcyg}
The largest deviation from a blackbody detected in the early spectra is around 810\,nm \citep{Smartt2017}, and is interpreted as the strong resonance lines at  1.0037, 1.0327, and 1.0915\,\(\mu\)m from Sr\(^+\) in the ejecta \citep{Watson2019}. These lines are modelled using the P~Cygni profile prescription described below, with the relative strengths of each of the lines set by the local thermodynamic equilibrium relation. A less prominent P~Cygni feature at 0.76$\mu$m likely associated with transition lines of Y\(^+\) has also been observed in the kilonova \citep{Sneppen2023b}.

The P~Cygni profile is characteristic of expanding envelopes where the same spectral line yields both an emission peak near the rest wavelength and a blueshifted absorption feature \citep{Jeffery1990}. The peak is formed  by true emission or by scattering into the line of sight, while the trough is due to absorption or scattering of photospheric photons out of the line of sight. As the latter is in the front of the ejecta this component is blueshifted. The P~Cygni profile is characterised by several properties of the KN atmosphere. The optical depth determines the strength of absorption and emission, while the velocity of the ejecta sets the wavelength of the absorption minimum. For this analysis we use the implementation\footnote{We adapt Ulrich Noebauer's \texttt{pcygni\_profile.py} \url{https://github.com/unoebauer/public-astro-tools} }
of the P~Cygni profile in the Elementary Supernova model \citep{Jeffery1990}, where the profile is expressed in terms of the rest wavelength $\lambda_0$, the line optical depth $\tau$, with velocity stratification parametrised with a scaling velocity $v_{\rm e}$, a photospheric velocity $v_{\rm ph}$, and a maximum ejecta velocity $v_{\rm max}$. Furthermore, we include a parameter describing the enhancement or suppression of the P~Cygni emission to improve the fit to the line shape. This parameter does not impact our constraints as it is predominantly the absorption component that provides the velocity constraints.   While this parametrisation assumes an optical depth with an exponential decay in velocity (i.e.\ $\tau(v) = \tau_0 \cdot e^{-v/v_{\rm e}}$; \citealt{Thomas2011}), we note that we can equivalently assume a power-law decay with the resulting constraints being indistinguishable.
This P~Cygni fitting framework follows the convention already established in \citet{Watson2019} and \citet{Sneppen2023} for fitting AT2017gfo.

To account for ejecta asphericity, we include in our P~Cygni profile prescription an additional and free-to-vary eccentricity $e$ and viewing angle $\theta_{\rm inc}$. This allows an independent probe of the sphericity by fitting the line shape to an expanding ellipsoidal photosphere. While in general constraints from the shape of spectral lines on the asymmetry of ejecta is degenerate with the viewing angle \citep{Hoeflich1996}, in this case the radio jet, the precision astrometry, and the gravitational standard siren in conjunction provide strong priors on the inclination angle of the merger, with \(\theta_{\rm inc} = 22\degree\pm3\degree\) \citep{Mooley2018,Mooley2022}. With a strong prior on the inclination angle, we can fit the  line shape to constrain the asphericity of the ellipse, ${v_{\perp}}/{v_\parallel}$. We find  $1.01\pm0.01$ and $0.995\pm0.006$ for epochs 1 and 2, respectively \citep{Sneppen2023}. We note that this uncertainty is only the statistical uncertainty of fitting the model to the data. Systematic effects such as blending with other possible (as yet unidentified) lines or time-delay and/or reverberation effects may shift these constraints. However, we argue that these systematic effects must be relatively small given the consistency in distance across statistically independent epochs as elaborated in Sect.~\ref{sec:sph}. 

\subsection{The expanding photosphere method}\label{sec:epm}
The expanding photosphere method (EPM) is a tool to measure the luminosity distance $D_L$ to objects with large amounts of ejected material \citep{Baade1926,Wesselink1946,Kirshner1974}. While the methodology was first developed for core-collapse SNe, \cite{Sneppen2023} demonstrated its applicability to KNe. The EPM assumes a simplified model of the ejecta as a photosphere in homologous and spherically symmetric expansion, so the photosphere size follows directly from the expansion velocity, measured from a Doppler line shift, and the time since explosion: $R_{\rm ph} = v_{\rm ph} (t-t_{\rm e})$. In KNe homologous expansion is expected to set in rapidly: less than a second after the merger for dynamical ejecta \citep{Rosswog2014} and up to $10^1-10^2$s for the post-merger ejecta \citep{Kasen2015}. There is now also good evidence that the outer ejecta of AT2017gfo was spherical \citep{Sneppen2023}. However, the sphericity assumption can be relaxed if the symmetry of the explosion can be quantified. Modelling and propagating these symmetry constraints yields a cross-sectional radius of the form 
\begin{equation}
    R_{\rm \perp} = v_{\rm \perp} (t-t_{\rm e}) = v_{\rm ph} (t-t_{\rm e}) \left[\frac{v_{\rm \perp}}{v_{\rm ph}}\right]
    \label{eq:R_perp}
,\end{equation}
where the ratio in the square bracket and $v_{\rm ph}$ come directly from the line Doppler shift and line shape constraints. 

The luminosity distance can be estimated by comparing the luminosity of the blackbody to the observed dereddened flux. The specific total luminosity inferred from observations is $L_{\lambda}^{\rm obs} = 4 \pi D_L^2 F_\lambda$, where \(F_\lambda\) is the wavelength specific flux, while the overall luminosity of a blackbody from Eq.~\ref{eq:Lumino} is \(L_{\lambda}^{\rm BB} = 4 \pi R_{\rm \perp}^2 \pi B(\lambda,T_{\rm eff}) f(\beta)\). Combining these with Equation~\ref{eq:R_perp}, the luminosity distance is thus

\begin{equation}
    D_L = v_{\rm ph} (t-t_{\rm e}) \left[\frac{v_{\rm \perp}}{v_{\rm ph}}\right] \sqrt{\frac{\pi B(\lambda,T_{\rm eff}) f(\beta)}{F_\lambda } }
    \label{eq:D_l}
.\end{equation}

Crucially, this estimate requires no calibration with known distances and is entirely independent of the cosmic distance ladder. Furthermore, for KNe the precise explosion time ($t_{\rm e}$) is provided by the gravitational wave signal. This ensures that each spectrum can yield an estimate of distance that is statistically independent from every other epoch, thereby testing the internal consistency and robustness of this EPM framework. As mentioned above, the blackbody colour and luminosity temperatures are likely to be very close (i.e.\ the dilution-factor correction is negligible), and the factor $f(\beta)$ can in principle be calculated analytically. There are therefore no major unknown quantities in the equation and each measured quantity (flux, velocity, time, and temperature)  is well constrained.

\begin{figure}
    \includegraphics[width=\columnwidth,viewport=24 24 555 490,clip=]{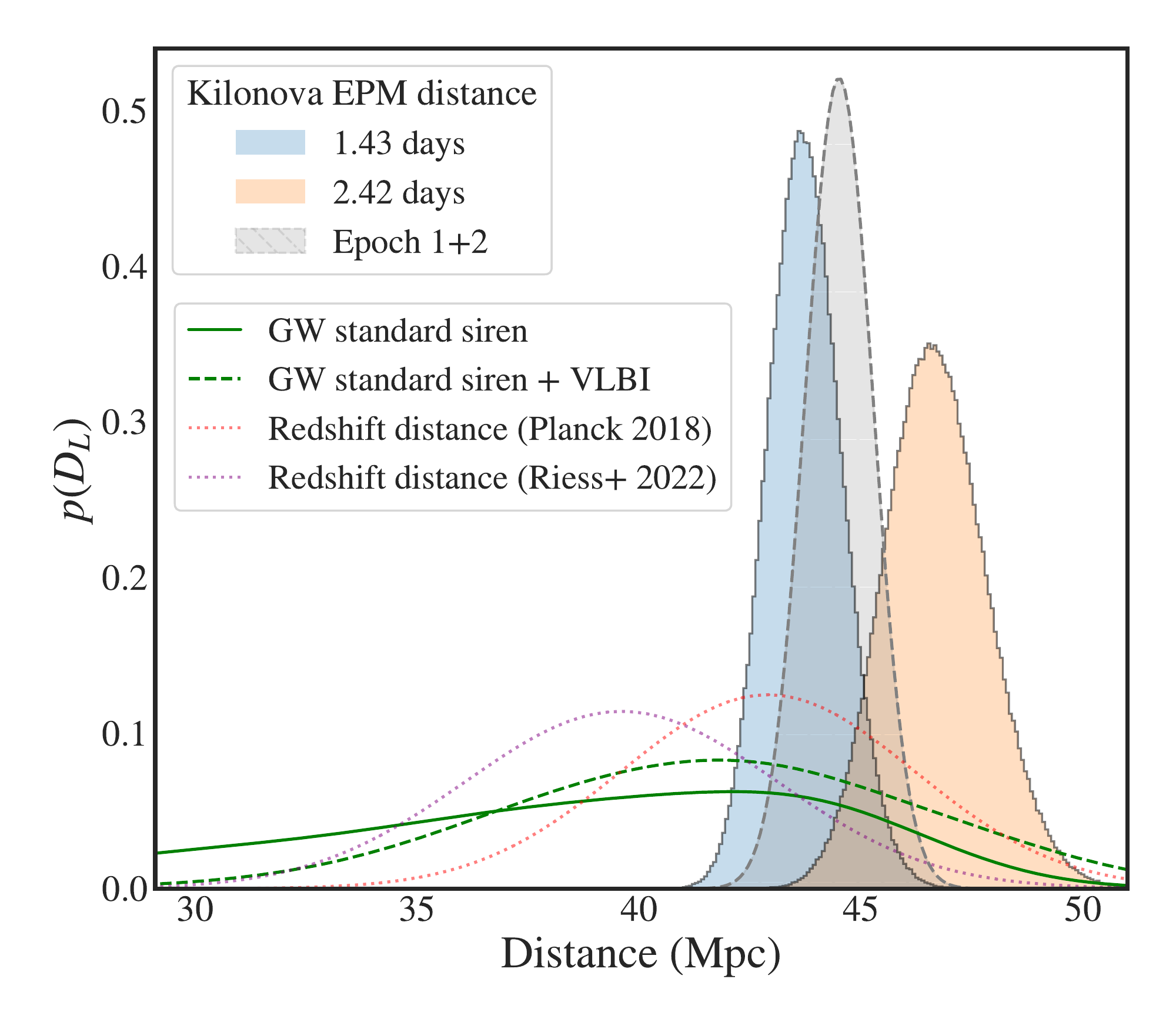}
    \caption{Posterior probability distributions of luminosity distances for each epoch sampled with MCMC runs. The distance estimate from the standard siren with and without very long baseline interferometry radio inclination angle data \citep{Mooley2018,Hotokezaka2019} are shown as dashed and solid green lines, respectively. Cosmological redshift distances are also plotted for \(H_0=67.36\pm0.54\) \citep{Planck2018} (dotted red) and \(H_0=73.03\pm1.04\) \citep{Riess2021} (dotted purple).} 
    \label{fig:d_l}
\end{figure}

Comparing the EPM estimated distance to the KN with the cosmological redshift of the host galaxy, we can now determine an independent constraint on the Hubble constant. To good approximation for $z \ll 1$, Hubble’s law gives the luminosity distance \citep[with $q_0 = -0.53$ for the Planck cosmological model;][]{Planck2018}:
\begin{equation}
    H_0 \approx \frac{c z_{\rm cosmic}}{D_L} \left( 1+\frac{1-q_0}{2}z_{\rm cosmic} \right)
    \label{dl_to_h0}
.\end{equation}Here $z_{\rm cosmic}$ is the recession velocity corresponding to pure Hubble flow (i.e.\ correcting for the peculiar velocity due to large-scale structure,  
$z_{\rm cosmic} = z_{\rm CMB} - z_{\rm pec}$).

\section{Application to AT2017gfo}
\subsection{Estimating the luminosity distance}\label{sec:dl}
We fit the spectra with a Planck function and a P~Cygni line profile associated with the Sr\(^+\) triplet at 1.0037, 1.0327, and 1.0915\,\(\mu\)m, to measure both the blackbody flux and the velocity of expansion. The X-shooter spectrum and the best-fit models for each of the first two epochs are illustrated in Fig.~\ref{fig:X-shooter-spec-4}.  Despite the simplicity of the methodology, the analysis provides a good fit to the data, consistent over multiple epochs. Using the EPM framework we deduce the luminosity distances for each epoch independently, with the posterior probability distributions shown in Fig.~\ref{fig:d_l}. This illustrates a few things of note. First, in each epoch the posterior distribution remains uni-modal with a single maximum in the likelihood landscape. Second, the statistical uncertainty in each epoch is small, at the level of 2--3\%. However, the difference between the distance measure for each epoch, which are statistically independent, is larger than this, at the level of about 5\% (i.e.\ comparable to the peculiar velocity uncertainty). This corresponds to a 1.9$\sigma$ discrepancy, which may indicate that the uncertainty on the distances is somewhat underestimated. The most obvious potential culprits in underestimating the uncertainty are the tight sphericity constraints from the line shape fitting and the dust extinction due to the host galaxy. We discuss the sphericity constraints in Sect.~\ref{sec:sph}, and the dust correction in Sect.~\ref{sec:dust}  

The estimated luminosity distance for the first and second epochs yield respectively $43.7\pm 0.9$\,Mpc and $46.6\pm 1.2$\,Mpc, which yields a joint constraint from epoch 1+2 of $44.5 \pm 0.8$\,Mpc. The unweighted average of these measurements yields $45.2 \pm 1.5$\,Mpc, which would suggest the fractional distance uncertainty is of the  order  of a few percent. These EPM distances are consistent with earlier distance estimates to the host galaxy NGC\,4993, including from surface brightness fluctuations \citep{Cantiello2018}, the fundamental plane \citep{Hjorth2017}, GW standard siren measurements \citep{Abbott2017a}, and combining the GW standard siren with constraints on inclination angle from very-long-baseline interferometry \citep{Hotokezaka2019}. 


\subsection{Constraints on $H_0$}\label{sec:h0} 
The recession velocity of the host-galaxy group is well constrained: $z_{\rm CMB} = 0.01110 \pm 0.00024$ \citep{Abbott2017a}. However, the contribution of the uncertainty on the peculiar velocity to $z_{\rm cosmic}$ is large due to the low redshift of the host galaxy NGC\,4993. Several estimates of $z_{\rm pec}$ have been made \citep{Hjorth2017,Howlett2020, Nicolaou2020}. The most recent analysis, using a reconstruction of observed galaxy distributions from a full forward modelling of large-scale structure 
and large-scale velocity flow with with Bayesian Origins Reconstruction from Galaxies  \citep[BORG;][]{Jasche2019} model, determines $z_{\rm pec} = 0.00124 \pm 0.00043$ \citep{Mukherjee2021}. We note that the peculiar velocity of the binary system relative to the host-galaxy is unimportant for any constraints as long as the binary is relatively close to the host or resides within the host. The binary itself is unlikely to have escaped its host galaxy given  the observed proximity and  the massive nature of the host (with comparable sGRB progenitors being bound within the tens of kiloparsec-scale) \cite{Fong2013}. 

Correcting the observed recession velocity with this peculiar velocity estimate yields a cosmic recession velocity of $z_{\rm cosmic} = 0.00986 \pm 0.00049$. This corresponds to a cosmological distance of $40.9 \pm 2.1$\,Mpc  given $H_0$ from the SH0ES measurement \citep{Riess2021} and $44.2 \pm 2.3$\,Mpc from Planck constraints \citep{Planck2018}. 
With a constraint on the luminosity distance of the order of  1–2\%, the peculiar velocity uncertainty of the order of 5\% is and will likely remain the dominant uncertainty in estimating $H_0$ using AT2017gfo. With future kilonovae, especially at somewhat higher redshift where the relative impact of the peculiar velocity is not as strong, this approach may yield more precise constraints on cosmology. A sample of KN hosts will also average out the effect of individual peculiar velocities. We discuss this possibility further in Sect.~\ref{sec:future}.

Given Equation~\ref{dl_to_h0} we determine the Hubble constant to be $H_0= 68.2 \pm 3.7$\,km\,s$^{-1}$\,Mpc$^{-1}$ and $H_0= 63.9 \pm 3.6$\,km\,s$^{-1}$\,Mpc$^{-1}$ for each of the two first periods and $H_0= 67.0 \pm 3.6$\,km\,s$^{-1}$\,Mpc$^{-1}$ from the joint constraints of epoch 1+2
. This is statistically consistent with the Hubble constant inferred from Planck \citep{Planck2018} and is within $2 \sigma$ of the local $H_0$ determination based on type\,Ia supernovae with Cepheid calibration \citep{Riess2021}. We address the effects of systematics in the next section. However, once again, we reiterate that these values of distance and Hubble constant should be taken as only indicative until full 2D or 3D radiative transfer modelling is done with reverberation effects and reasonably complete line and opacity effects considered.  We consider the limitations of our ansatz of an ellipsoidal blackbody below.


\section{Potential systematics}\label{sec:sys}
We   show above that the EPM application to KNe can provide precision cosmological distances and independent constraints on $H_0$. We discuss now its assumptions and systematics to gauge their potential impact.

\begin{figure}
    \centering
    \includegraphics[width=\linewidth, viewport=18 22 460 380,clip=]{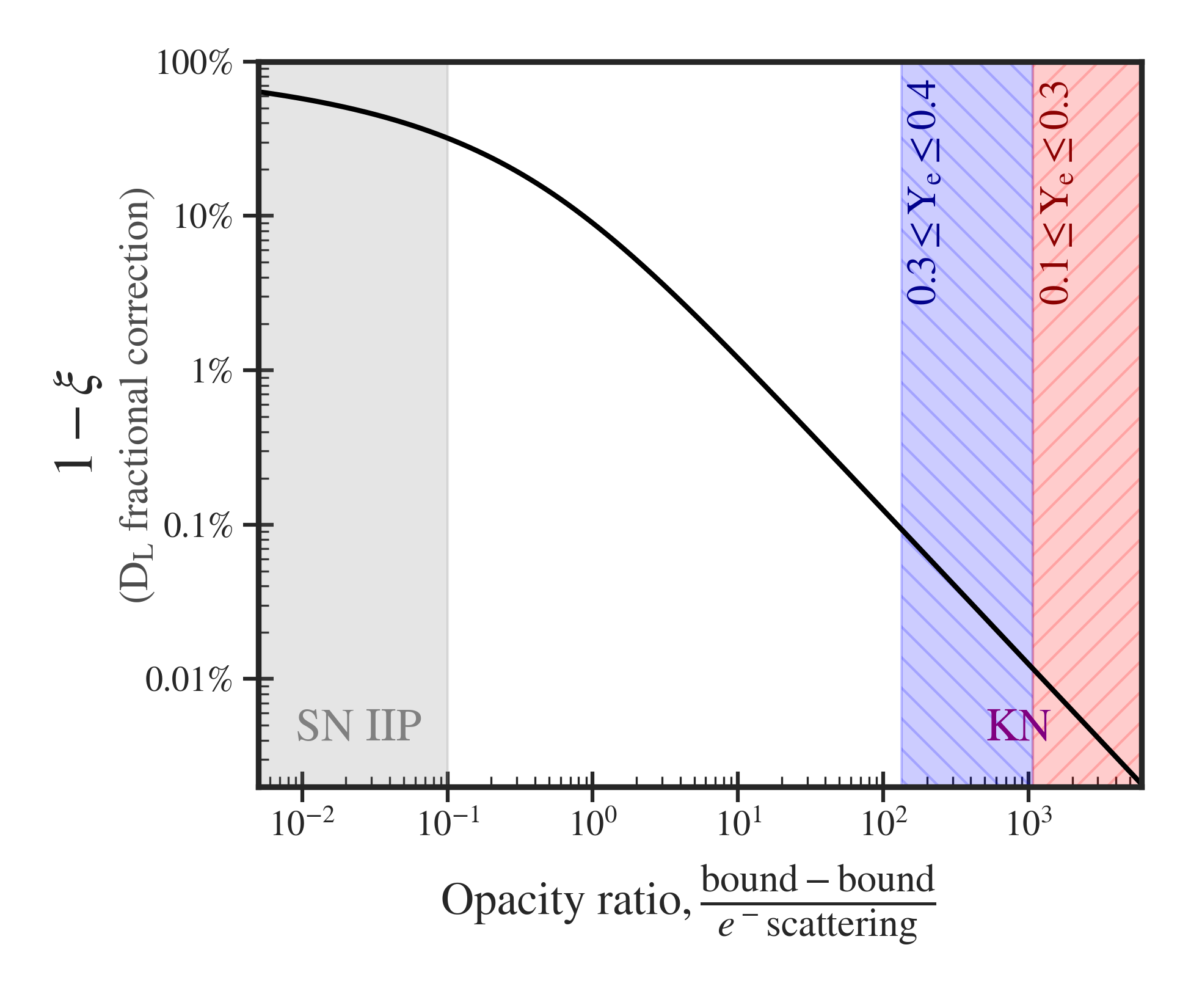}
    \caption{
    Fractional correction to the estimated luminosity distance from dilution factors, $1-\xi$, as a function of   opacity ratio, $\kappa_{\rm{b-b}}/\kappa_{e^{-} \rm{scat}}$. As the opacity ratio is a function of wavelength,  we report the integrated Planck mean opacity. 
    When electron-scattering dominates, as in the case of SN\,IIP \citep{Sim2017}, the photospheric radius inferred from lines will be larger than the thermalisation radius of the blackbody inferred from its flux, thus requiring the introduction of a significant dilution-factor (grey region). However, for neutron-star mergers the bound-bound opacity dominates over electron-scattering, regardless of whether one uses opacities based on lists of known lines \citep[e.g.][]{Tanaka2013} or the higher opacities used in this figure, which are based on more complete theoretically-calculated line-lists \citep{Tanaka2020}. Such dominance of the bound-bound opacity implies negligible corrections to the distance (blue and red hatched region).} 
    \label{fig:dilution}
\end{figure}

\subsection{Dilution factors}
Historically EPM analyses of SNe required correcting the luminosity obtained from specific photometric bands by multiplying by a factor $\xi^2$. Here $\xi$ was originally introduced because the large contribution of electron-scattering opacity to the total opacity within the photospheres of SN envelopes dilutes the continuum flux relative to what would be inferred for a blackbody continuum with radius inferred from the line velocity \citep{Hershkowitz1986}. However, in practice dilution factors also account for any deviation from the blackbody including the presence of lines \citep{Vogl2019}. These dilution factors are determined with spectral synthesis modelling, but with large discrepancies between competing models, this constitutes one of the most significant sources of uncertainty in the application of EPM to SNe IIP \citep{Eastman1996,Dessart2005,Dessart2015}. For SN IIP, the electron scattering opacity, $\kappa_{e^{-} scat}$, is around 10--100 times greater than the bound-bound opacity, $\kappa_{b-b}$, in the optical, and 100--1000 times greater in the infrared \citep{Sim2017}.  Thus, for SN IIP environments the ratio of bound-bound to electron scattering opacity is low for the wavelengths in this analysis. In Fig. \ref{fig:dilution}, the grey region indicates this regime, where electron-scattering dominates which produces a significant dilution factor. The dilution factors dependence on the opacity ratio (shown with the the black line) is derived from a simplified model, which assumes a plane parallel scattering atmosphere, a Planck function varying linearly with optical depth and a constant ratio of the opacity contributors with depth (e.g. Sect. 4.1 in \citealt{Dessart2005}). 


In contrast, for KNe the environment dilution factors are minor. First, the absorption and emission features of the spectrum are explicitly  parametrised and fit. Second, the contribution of electron-scattering opacity to the total opacity is expected to be negligible in KN atmospheres, as $\kappa_{b-b}$ is typically several orders of magnitude higher in the UV-optical \citep{Kasen2013}, while the  number of free electrons per unit mass is reduced by $\approx 10^{2}$ relative to ionised hydrogen. Thus, electron scattering opacity is expected to be less than the \rprocess line opacity at the relevant wavelengths, and so the dilution effect should be very small, which results in a tight correspondence between thermalisation and photospheric radii. In \cite{Tanaka2020} the Planck mean opacity from bound-bound transitions (at the characteristic temperatures analysed here) ranges from $0.4$ to $40\, \rm{cm^2 g^{-1}}$ for models that are respectively light or heavy \rprocess dominated. These mean opacities are 130 and 13\,000 times larger than the corresponding electron-scattering opacity for NSM ejecta, $\kappa_{e^{-} scat}\approx3\cdot10^{-3}\ \rm{cm^2 g^{-1}}$ \citep[e.g. Eq. 12 in][]{Kasen2013} and are indicated in the blue and red hatched regions in Fig. \ref{fig:dilution}. Regardless of the exact \rprocess elementary abundances, dilution-factors are expected to be negligible in KNe.


Another line of reasoning, which illustrates the smaller importance of dilution factors in KNe is the consistent measurements of distance across epochs. When dilution-factors are required to correctly fit the distance such as is the case of SN IIP, the invoked dilution-corrections have to vary across epochs. This is because $\kappa_{e^{-} scat}$ decreases with time (because the temperatures and ionisation-levels decrease), while $\kappa_{b-b}$ increases over photospheric epochs in models \citep{Tanaka2013}. For example, the ratio $\kappa_{\rm{b-b}}/\kappa_{e^{-} \rm{scat}}$ increases by a factor of 10 from 1.5 to 3.5 days post-merger for the models presented in \citet{Bulla2019}. However,  while the epochs examined here (in addition to epochs 3, 4, and 5  analysed in \citealt{Sneppen2023})  probe significantly different phases and characteristic opacities, they still yield consistent measurements of distance. The only physical regime where changing the ratio of opacities drastically does not lead to a shift in distance, is when the bound-bound opacities dominate and $\xi\approx1$.

\begin{figure*}[t]
    \includegraphics[width=\linewidth,viewport=5 25 920 500,clip=]{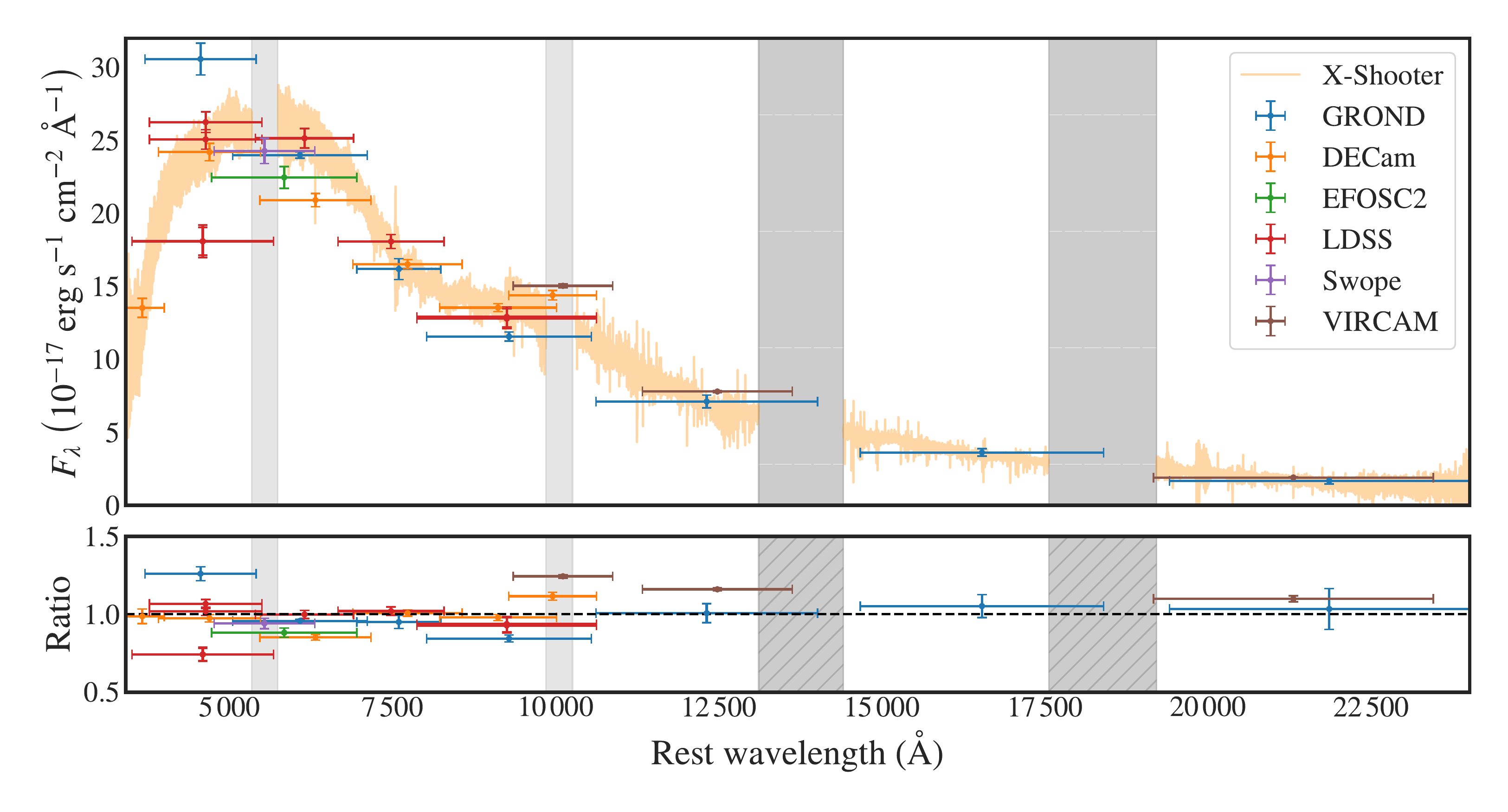}
    \caption{X-shooter spectrum and photometry for the KN AT2017gfo from UV--NIR for observations from 1.35 to 1.45 days after the NS merger GW170817. The dark shaded bars indicate telluric regions; the light shaded bars indicate overlapping noisy regions between the X-shooter UVB, VIS, and NIR arms. The ratio of photometric points to the spectrum (illustrated in the lower panel) shows that while there is significant scatter between photometric points, the normalisation of the spectrum is consistent with the average of the photometric fluxes to within 1\%. Data are taken from the literature as follows: GROND and EFOSC2 \citep{Smartt2017}; DECAM \citep{Cowperthwaite2017}; LDSS \citep{Shappee2017}; Swope  \citep{Coulter2017}; VIRCAM  \citep{Tanvir2017}.} 
    \label{fig:calib}
\end{figure*}

\begin{figure*}
    \includegraphics[width=\linewidth, viewport=20 22 720 305, clip=]{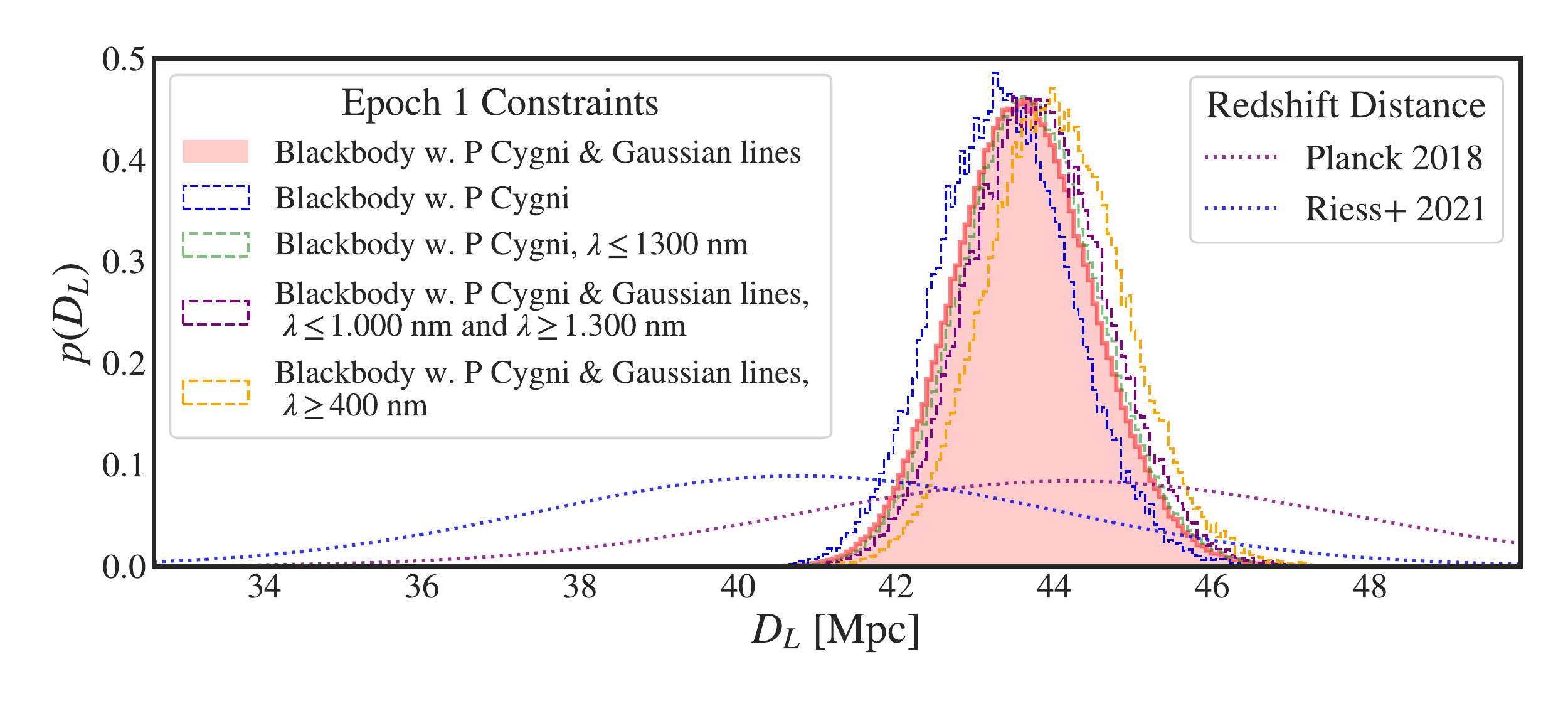}
    \caption{Posterior probability distributions of luminosity distances for the first epoch spectrum with and without modelling of different emission features. This shows the size of effects related to the choice of how the continuum and lines are modelled. The canonical model (red) is a blackbody continuum with Sr\(^+\) elementary supernova P~Cygni lines and Gaussian emission lines at 1.5\,$\mu$m and 2\,$\mu$m. The green histogram shows the effect of limiting the analysis to wavelengths below 1300\,nm and using only a blackbody continuum with Sr\(^+\) P~Cygni, which avoids the need to include the 1.5\,$\mu$m and 2\,$\mu$m lines in the model. Very similar distances are found fitting the full dataset with only a blackbody and the Sr$^+$ P~Cygni lines (blue) or only fitting the absorption wavelengths of the P~Cygni (purple) (i.e.\ excluding the P~Cygni emission wavelengths, $1\,\mu\mathrm{m} < \lambda < 1.3\,\mu\mathrm{m}$, from the fit). Lastly, the distance is robust to the lower wavelength cutoff as seen by fitting the canonical model to wavelengths $ \lambda > 330$\,nm (red) or $\lambda > 400$\,nm (yellow). The fitting result is therefore very robust to the details of how the fit is performed and what data are included. Distances inferred from the cosmological redshift are also plotted for \(H_0=67.36\pm0.54\) \citep{Planck2018} (dotted purple) and \(H_0=73.03\pm1.04\) \citep{Riess2021} (dotted blue).
    } 
    \label{fig:variety}
\end{figure*}

\subsection{The blackbody assumption}
It is worth noting that the observed blackbody of the spectrum is difficult to reproduce and explain within current radiative transfer simulations. This is due to the complexities associated with a line-dominated opacity and problems in modelling r-process elements, which have incomplete atomic data and line lists \citep{Gaigalas2019,Tanaka2020,Gillanders2021}. For now, therefore, the blackbody origin cannot be derived from first principles, and is instead an empirical assumption based on the close similarity to a blackbody in both luminosity and spectral shape at all wavelengths from the UV to the NIR in the early spectra of AT2017gfo. Therefore, this methodology relies on the ansatz that the emitting plasma really is a blackbody to better than the required accuracy (currently a few percent in the luminosity). This ansatz will be tested in the coming years given ongoing work in improving the relevant atomic data and radiative transfer in kilonova atmospheres \citep[e.g.][]{Domoto2021,Gaigalas2019,Gillanders2022,Flors2023}. However, the blackbody assumption is not essential to the method, and, as in the spectral-fitting expanding photospheres method \citep[SEAM,][]{Baron2004}, a complete  radiative transfer simulation should ultimately be used.

Given the significant light travel time-delays, the rapidly cooling surface and the varying projected velocity of the relativistic Doppler corrections, we note that the different latitudinal parts of the surface are observed with different effective temperatures. This results in a multi-temperature blackbody (i.e.\ a blackbody convolved over a range of different effective temperatures). However, as shown in \cite{Sneppen2023_bb} for the characteristic velocities and cooling-rates of AT2017gfo these effects cancel and the variations with wavelength remain below 1\% over the entire spectral range, suggesting that any angular dependence over the observed surface is negligible.

Even so, the EPM framework can straightforwardly be extended with a modified blackbody framework given any prescribed relation between temperature and time. Using the predicted power-law decay from a large ensemble of \rprocess isotopes \citep{Metzger2010,Wu2019b}, which is also observed from the fading of the light curves for AT2017gfo \citep{Drout2017,Waxman2018}, the derived distances remain robust to the assumed cooling rate. In other words,  given the functional form 
\begin{equation}
T_{\rm emi}(t) = T_{\rm ref} \cdot \left(\frac{t}{t_{\rm ref}} \right)^{-\alpha}
,\end{equation}
the difference in inferred distance between assuming a constant temperature ($\alpha=0$) and the power-law decay inferred from cross-spectra comparisons ($\alpha\approx0.4-0.5$) are below 1\% \citep{Sneppen2023_bb}.

Furthermore, the wavelength-dependence of opacity and the possibility of a significant radial temperature gradient could imply that different wavelengths probe different radii, and thus temperatures. This radial convolution would again produce a modified blackbody (dependent on the density profile, temperature gradients, and wavelength-dependence of the opacity), which would be broader than a single-temperature blackbody spectrum. However, the spread of temperatures allowed observationally from fits to the kilonova spectra, as calculated in \citet{Sneppen2023_bb}, suggests this must be a rather limited effect. As a check on the wavelength-dependence of the blackbody, we fit the UV arm and optical--NIR arms separately in epochs 1 and 2. These yield blackbody temperatures consistent at the level of 2\% and 4\%, respectively, indicating a consistent single temperature at least from the blue into the near-infrared.

\subsection{Flux calibration}
The flux calibration of the X-shooter spectra is   consistent with coincident photometry from GROND, DECam, EFOSC2, LDSS, Swope, and VIRCAM, as illustrated for Epoch~1 in Fig.~\ref{fig:calib}. Therefore we follow \citet{Watson2019} and \citet{Sneppen2023} and do not artificially scale the spectra to the available photometric data, which have significant internal scatter. 

Nevertheless, we do use the photometric data to ensure our errors are robust. We compute the standard error of the ratio of the photometric fluxes to the synthetic flux from the spectrum integrated over the filter profile for all photometric points (including all observations from GROND, DECam, EFOSC2, LDSS, Swope, and VIRCAM). We add this standard error to the normalisation uncertainty from the fit. In practice, this is implemented by taking the posterior distribution of the normalisation sampled with MCMC and multiplying each instance of the sampling with a random number drawn from a Gaussian distribution with a mean of 1 (as the photometry and spectrum are consistent) and a standard deviation equal to the standard error of the flux calibration. The standard error of the photometric points relative to the X-shooter spectra are 2.5\% and 3.8\% for epoch~1 and epoch~2, respectively. This method is very cautious; in \citet{Sneppen2023} the errors were increased by the uncertainty on the weighted average between the photometric points and the X-Shooter spectrograph, which adds an additional uncertainty of 0.5\% and 0.8\% on the normalisation for epochs~1 and 2, respectively. For this analysis, we chose the unweighted standard error as coincident photometric points at comparable wavelengths show a low level of discrepancy between different telescopes (see Fig.~\ref{fig:calib}), which suggests that the stated uncertainties in the available photometry may be underestimated. 

\subsection{Dust extinction}\label{sec:dust}

Dust extinction is one of the major uncertainties in SN cosmology and may yet hold some clues to the Hubble tension \citep[e.g.][]{Wojtak2022}. For kilonova distance measurement, we must therefore also carefully consider the level and uncertainty of the extinction. The extinction originates   in the Milky Way and potentially in the host galaxy. In the case of AT2017gfo, the Milky Way dust dominates.

\subsubsection{Milky Way extinction}\label{sec:MWdust}
In the direction of the host galaxy, the Galactic reddening estimate is $E(B-V) = 0.1053 \pm 0.0012$ based on dust emission data from the \emph{IRAS} and \emph{COBE} satellites, calibrated to an extinction measure using SDSS stellar spectra \citep[called the SFD maps;][]{Schlegel1998,Schlafly2011}. This value is consistent with estimates based on Galactic Na\,D absorption lines \citep{Shappee2017,Poznanski2012}, though the Na\,D correlation with dust extinction has large scatter \citep{Poznanski2011} and is not a useful constraint in this context. The $\sim1$\% uncertainty in $E(B-V)$ quoted above is very small and is an underestimate of the total absolute extinction uncertainty. We address this further here. The dust maps \citep{Planck2016} calculated using data from the \emph{Planck} satellite, disentangled from cosmic infrared background anisotropies using the generalized needlet internal linear combination (GNILC) method, and combined with the SFD maps and IRIS maps \citep{Miville-Deschenes2005} is in agreement with the above numbers: the \emph{Planck}-GNILC data suggest $E(B-V) = 0.1068 \pm 0.005$ along this sightline. The 3D dust maps based on extinction measurements using stellar photometry from stellar photometry of 800 million stars from Pan-STARRS 1 and 2MASS \citep{Green2018} independently gives $E(B-V) = 0.116 \pm 0.007$ in this direction. None of these Galactic extinction uncertainties would substantially change the uncertainty on the distance (see below) and all these estimates are consistent with each other at the level of 10\%. We use $E(B-V)=0.11$ which is the weighted average of \cite{Green2018} and \cite{Planck2016}, and conservatively use a 15\% uncertainty including the absolute uncertainty in this direction. The extinction correction is implemented as laid out in \citet{Watson2019}, which assumes the \citet{Fitzpatrick1999,Fitzpatrick2004} reddening law with $R_V = 3.1$. In addition, we also adopt a 10\% uncertainty on the $R_V$, which we propagate throughout the uncertainty estimates of this analysis. This is based on the standard deviation of $R_V$ values found along different Milky Way sightlines \citep{Fitzpatrick2007,Mortsell2013}.

\subsubsection{Host galaxy extinction}\label{sec:host_dust}
To get an estimate of the likely level of extinction in the host galaxy along this sightline, we estimate the dust surface density at the location of AT2017gfo in its host galaxy and convert it to an extinction estimate as follows. We start with the stellar mass surface density at 2.1\,kpc based on the S\'ersic model fit by \citet{Blanchard2017}, and convert it to a dust surface density using the typical stellar mass-to-dust mass ratio for elliptical galaxies measured by the \emph{Herschel} Reference Survey, \(\log(M_D/M_*) = -5.1 \pm 0.1\) \citep{Smith2012}. This gives us a surface dust density of 2780\,M\(_\sun\)\,kpc\(^{-2}\). Assuming AT2017gfo is in the middle of this dust, we divide by two. We then convert this to an extinction along this line of sight using the gas-to-dust mass ratio for the Milky Way, \ 90--150 \citep{Magdis2012,Galametz2011,Draine2007}, and the relation between gas column density and extinction, \(N_H \sim 2\times10^{21}\,\mathrm{cm}^{-2}\,A_V\) \citep{Watson2011,Bohlin1978}. This returns a value of \(A_V \simeq 0.01\). Equally, if we use the conversion from \citet{PlanckCollaboration2016} directly from the dust surface density, \(\Sigma_{M_d}\) (i.e.\ \(0.74 \times10^{-5} \times \Sigma_{M_d} M_\sun\,\mathrm{kpc}^{-2}\)) we get a similar result. Our estimate is highly uncertain, but small (about 3\% of the estimated Milky Way extinction; the redshift effect on the dust extinction curve is negligible at this distance),  and also small compared to our uncertainty on the Milky Way extinction, so following \citet{Drout2017} and \citet{Watson2019}, we do not correct for reddening from dust within the host galaxy. The lack of observed Na\,D features in the spectrum at the redshift the host galaxy \cite{Shappee2017} also favors quite a low intrinsic absorption within the host \citep[$E(B-V)\lesssim0.03$ using the relation of][]{Poznanski2012}. \citet{Blanchard2017} find a similar upper limit for the total dust attenuation of the light of NGC\,4993 (\(A_V<0.11\)\,mag) based on modelling the spectral energy distribution of the whole galaxy, again, consistent with our low estimate. While in this case we find the host galaxy extinction negligible compared to the Milky Way extinction, a better estimate would be useful.

\subsubsection{Effect of extinction on the distance estimate}\label{sec:dust_distance}
There are two opposing effects on the inferred luminosity distance if a significant amount of dust is not accounted for. First, attenuation of the light  would make the object appear fainter, which would lead us to infer a greater distance. Second, the reddening effect of the dust extinction would make the observed blackbody appear cooler (i.e.\ the kilonova would be inferred from the fit to be intrinsically less luminous, and thus closer to the observer). These two effects largely counteract each other, with the apparent cooling effect being slightly dominant for the wavelengths and temperatures of AT2017gfo at the epochs analysed here. Thus, the competing effects of biasing the apparent temperature and luminosity mean that the uncertainty in the dust extinction level we infer here only corresponds to around a 1.5\% fractional uncertainty in the distance.

Overall, finding a more accurate way to measure host galaxy extinction for future events may be important, at least in some cases, if we are to use kilonovae to estimate \(H_0\) to high precision. In general, however, kilonovae occur in less dusty environments than SNe, so the extinction corrections will typically be smaller than for most SNe. Furthermore, it is likely that on average, future KNe will have lower Galactic extinction, with the median Galactic extinction values from X-ray-selected GRB samples being around $A_V = 0.1-0.2$ \citep[e.g.][]{Watson2011}.


\subsection{Evolving spectral line emission}
While the earliest epochs of AT2017gfo are well approximated by a blackbody spectrum \citep{Malesani2017,Drout2017,Shappee2017}, over time, the spectrum increasingly deviates from a blackbody with more numerous and progressively stronger spectral components. Broad excess emission is observed at wavelengths of around $1.5\,\mu$m and $2\,\mu$m; there is an emission feature at $\sim1.4\,\mu$m within the telluric region, clearly observed in the \emph{Hubble Space Telescope} spectra from the fifth and tenth days post-merger \citep{Tanvir2017}. 
The derived constraints are robust at early times, whether one includes these components in a parametrised fashion, excludes the spectral regions containing these components, or indeed neglects modelling these features altogether (see Fig.~\ref{fig:variety}). Thus, while the modelling of later epochs may be affected by whether and how these features are accounted for, the constraints from the earliest epochs are little affected by these emerging spectral components. For the constraints presented in this paper, we include Gaussian emission lines as nuisance parameters to explicitly parametrise the observed emission components at $1.5\,\mu$m and $2\,\mu$m. 

Thus, in constraining the continuum we can use the broad wavelength coverage from the UV at 330\,nm to the NIR at 2250\,nm (excluding the telluric regions and noisy regions between UV, optical, and NIR arms). The analysis is not very sensitive to the specific wavelength range used (see Fig.~\ref{fig:variety}). Naturally, the constraining power decreases given a more limited wavelength range. The blue and visible flux is important for fitting the blackbody peak (and in extension the blackbody temperature and normalisation); however, for reasonable choices of the short wavelength cutoff (i.e.\ 330\,nm to 400\,nm) the derived distance remains highly consistent. 

\subsection{Line blending}
The identification of the $1\,\mu$m P~Cygni feature as being primarily due to Sr\(^+\) is consistently recovered in radiative transfer simulations \citep{Watson2019,Domoto2021,Gillanders2022} as 1) strontium is a highly abundant element in both the \rprocess abundance of the sun and in nucleosynthesis models (being near the first \rprocess peak), 2) Sr\(^+\) has three strong  lines at the observed rest-wavelength, and 3) the low-lying energy levels are easily accessible in local thermodynamic equilibrium for the characteristic temperatures in the kilonova atmosphere. However, a potential concern is that blending with other as-yet-unrecognised lines not in the available line lists could bias the inferred properties from the Sr\(^+\) P~Cygni profile. As discussed in \citet{Sneppen2023}, such a significant biasing line is not observationally required by the data and would be unlikely to produce the very consistent distances measured across multiple epochs. Furthermore, the velocity derived from Sr\(^+\) is consistent with the velocity stratification from the detection of a newly discovered yttrium P~Cygni feature at $\sim0.76\,\mu$m discovered in later epoch spectra \citep{Sneppen2023b}. Thus, any potential line blending bias would have to act coincidentally to reproduce consistent distance estimates and symmetry not merely across epochs, but also across different lines probing different wavelengths. 

To obtain a consistent distance across different epochs would require that the blending lines matched the Sr\(^+\) lines in strength over time, despite the major changes in density and temperature (as well as possible composition) that we   observe at each epoch. In addition, this requires the blending lines to mimic a Sr\(^+\) P~Cygni profile, coherently shifting in wavelength and receding deeper within the ejecta at exactly the rate predicted by the blackbody continuum. On the basis of this argument, it seems unlikely that line blending is substantially biasing the determined photospheric velocities, although in principle it cannot be excluded, and therefore remains a systematic uncertainty. This effect could be quantified with a full radiative transfer model with a complete atomic dataset for the relevant elements.

\subsection{Reverberation effects} 
A significant issue with current P~Cygni codes is the lack of reverberation effects which current models do not include. 
Reverberation effects happen because  the strength of the observed lines are dynamically evolving, and due to different light travel-times, the flux from different wavelengths in the line were emitted at different times. In conjunction, these imply that the constraints may be improved with a full 3D (or at least  2D) radiative transfer model. Modelling these higher order effects are outside the scope of the simplicity of our current framework and this initial analysis, but we note that reproducing kilonova spectra and their temporal evolution within an ensemble of self-consistent radiative transfer models could improve the confidence in, and the quality of, the constraints compared to these preliminary results. 


\begin{figure}
    \centering
    \includegraphics[width=\linewidth, viewport=38 40 550 480,clip=]{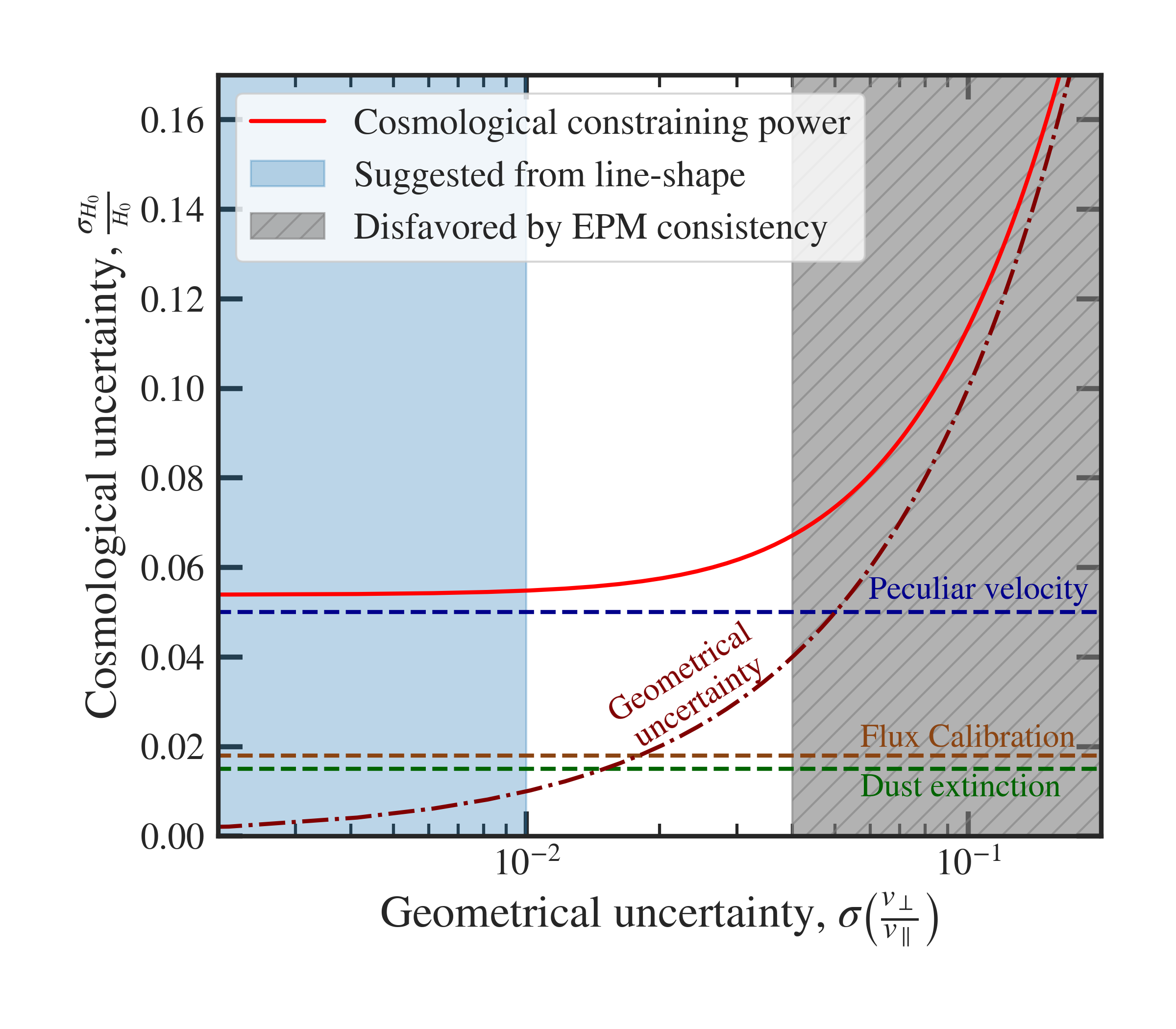}
    \caption{Cosmological constraints attainable for AT2017gfo as a function of the constraints on the ejecta geometry. The uncertainty associated with peculiar velocity modelling dominates for a well-constrained geometry, while a poorly constrained shape would dominate the error-budget when the fractional uncertainty exceeds 5\%. The small scatter across the  independent datasets from days 1.4--3.4 post-merger \citep{Sneppen2023} puts an upper bound on the statistical uncertainty on the geometry of a few percent (indicated with grey shading). Additionally, prior line shape constraints (blue shading) suggest percent-precision geometrical constraints may be attainable \citep{Sneppen2023}. Thus, both the EPM consistency and line shape analysis suggest, that AT2017gfo is in the regime where peculiar velocity modelling is the dominant uncertainty.
    } 
    \label{fig:Lineshape}
\end{figure}

\subsection{Measuring the sphericity}
\label{sec:sph}
A high degree of sphericity is indicated by  the line shape and  the consistently spherical measurement from the EPM framework across epochs. First, the characteristic fractional uncertainty on geometry from the line shape fitting is of the order of  1\%. Second, the 2\% spread in the statistically independent measurements of the sphericity index $\sigma_\Upsilon$ from epochs 1, 2, and 3 from the EPM analysis in \cite{Sneppen2023} provides an estimate of the upper bound on the internal uncertainty of this measurement. However, there is a slight evolution of both EPM and line shape constraints in the same direction and of comparable magnitude over time, hinting that the initially spherical ejecta geometry becomes slightly and increasingly prolate. While we account for this very mild monotonically increasing prolateness from the line shape, the majority of the observed scatter in distance estimates might still be caused by an underlying evolution in the geometry if it is not fully accounted for. That is because the EPM distance inferred from Eq.~\ref{eq:D_l} becomes slightly and monotonically larger in later epochs \citep{Sneppen2023} in the same way as the prolateness inferred from the line shape.


Additional constraints on the kilonova's asymmetry from spectropolarimetry at 1.46\,days are consistent with this spherical geometry, with the  observed low degree of polarisation being consistent with intrinsically unpolarised emission scattered by Galactic dust \citep{Covino2017,Bulla2019}. This indicates a symmetrical geometry of the emitting region,  with a very loose constraint on the viewing angle of 65\degr\ of the pole. While these measurements are not very constraining, spectropolarimetry could provide useful independent constraints of geometry and sphericity for future kilonovae.



For this analysis, we propagate the geometrical uncertainty from the line shape in \cite{Sneppen2023} into the final cosmological constraint. However, as noted, the geometrical constraints from line shapes may be affected by reverberation effects or blending lines, so these constraints should be considered preliminary to such improvements in radiative transfer modelling of KNe. Nevertheless, given the dominance of the peculiar velocity uncertainty, a somewhat looser constraint on sphericity would not drastically impact the $H_0$ error budget, as indicated in Fig.~\ref{fig:Lineshape}. It would require an increase of a factor of several  in the uncertainty in sphericity to make it comparable to the peculiar velocity uncertainty for AT2017gfo. Thus, despite the degeneracy in line shape fitting between geometry constraints and viewing inclination angle, moderately looser inclination constraints, such as that provided by using only the radio jet $\theta_{\rm inc}=21 \pm 7$ \citep{Mooley2018}, will not significantly increase the error budget. 
Inclination angle observations are currently necessary for strong line shape constraints. Unfortunately, a significant fraction of expected KNe may have no jet observations \citep{Colombo2022} we therefore cannot rely on them having well-constrained geometries. On the other hand, if a very high degree of sphericity proves to be a common feature of KNe, this may not be problematic. With improved numerical simulations, which can consistently explain and reproduce the observed geometry of future KNe, the deviation from sphericity and spread of geometry within such simulations can also be propagated within this framework. 

\subsection{Continuum relativistic and time-delay correction}\label{sec:continum}
The systematic effects and uncertainties discussed previously are important at the percent level for constraining distances, which for any $H_0$ constraints are well below the 5\% fractional uncertainty in modelling the peculiar velocity. Including the full relativistic and time-delay corrections on the continuum luminosity matters at the \(\sim10\%\) level for the characteristic velocity of the early kilonova ejecta. Thus, it is important to get this right, and in the following we discuss the time-delay and relativistic correction $f(\beta)$ used in Eq.~\ref{eq:Lumino}. 

First, the significant  time-delay of the mildly relativistic velocities induces a geometrical effect. In order to be observed at the same time, light arriving from the limb of the ejecta must have been emitted earlier than the light from the nearest front, at a time when the photosphere had a smaller surface area. As first derived in \cite{Rees1967}, this defines a surface of equal arrival time $r_{\rm ph}(\mu)/r_{\rm ph}(\mu=1) = (1-\beta)/(1-\beta \mu)$, where $\mu = \cos(\theta)$, with $\theta$ being the angle between expansion and the line of sight.

Second, the specific intensity is a blackbody \(B(\lambda',T')\) with temperature \(T'\) at wavelength \(\lambda'\) in the co-moving frame of the ejecta, which in the observed frame is inferred as a blackbody \(B(\lambda,T)\) with an observed temperature \(T\) and wavelength \(\lambda\). The relationship between emitted and observed temperature $T(\mu) = \delta(\mu) T'$ is given by the relativistic Doppler correction \(\delta(\mu) = \left(\Gamma(1-\beta\mu)\right)^{-1}\) \citep{Ghisellini2013}. Here the Lorentz-factor $\Gamma$ represents the relativistic time dilation, while the latter is the Doppler correction, which depends on the projected velocity along the line of sight $\mu$. We note that this means that it is a mathematical property of a blackbody in bulk motion, and that while the relativistic transformations boost the observed luminosity, this is exactly matched by a higher observed temperature. %

However, as the ejecta's projected velocity varies across different surface elements (from head-on expansion, $\mu = 1$, to orthogonal, $\mu=0$), so too will the Doppler correction   vary. This results in a second-order correction, where a convolution of relativistic Doppler corrections incoherently shift the spectrum. The deviation generally depends on the spectral shape, but for this analysis the variations with wavelength are below 1\% over the entire spectral range \citep{Sneppen2023_bb}. Thus, for the mildly relativistic velocities of relevance here, the luminosity-weighted spectrum remains largely well described by a single blackbody \(B(\lambda,T_{\rm eff})\), so the relativistic correction is not central for the cosmological constraints. 

In conjunction, these transformations yield an observed luminosity \citep{Sneppen2023}
\begin{align*}
    L_{\lambda}^{\rm BB} &=  \int I_\lambda \mu \ d\mu \ d\phi \ dA \\
    &= 2 \pi \int_\beta^1 \delta(\mu)^5 B(\lambda', T') \left(4 \pi R_{\rm ph}^2 \left(\frac{1-\beta}{1-\beta \mu}\right)^2 \right) \mu d\mu  \\
    &= 4 \pi R_{\rm ph}^2 \left[2 \pi \int_\beta^1 B(\lambda, T(\mu)) \left(\frac{1-\beta}{1-\beta \mu}\right)^2 \mu d\mu \right] \\
    &= 4 \pi R_{\rm ph}^2 \left[f(\beta)\;B(\lambda,T_{\rm eff})\right],
\end{align*}
\begin{figure}
    \centering
    \includegraphics[width=\linewidth,viewport=1 5 503 382,clip=]{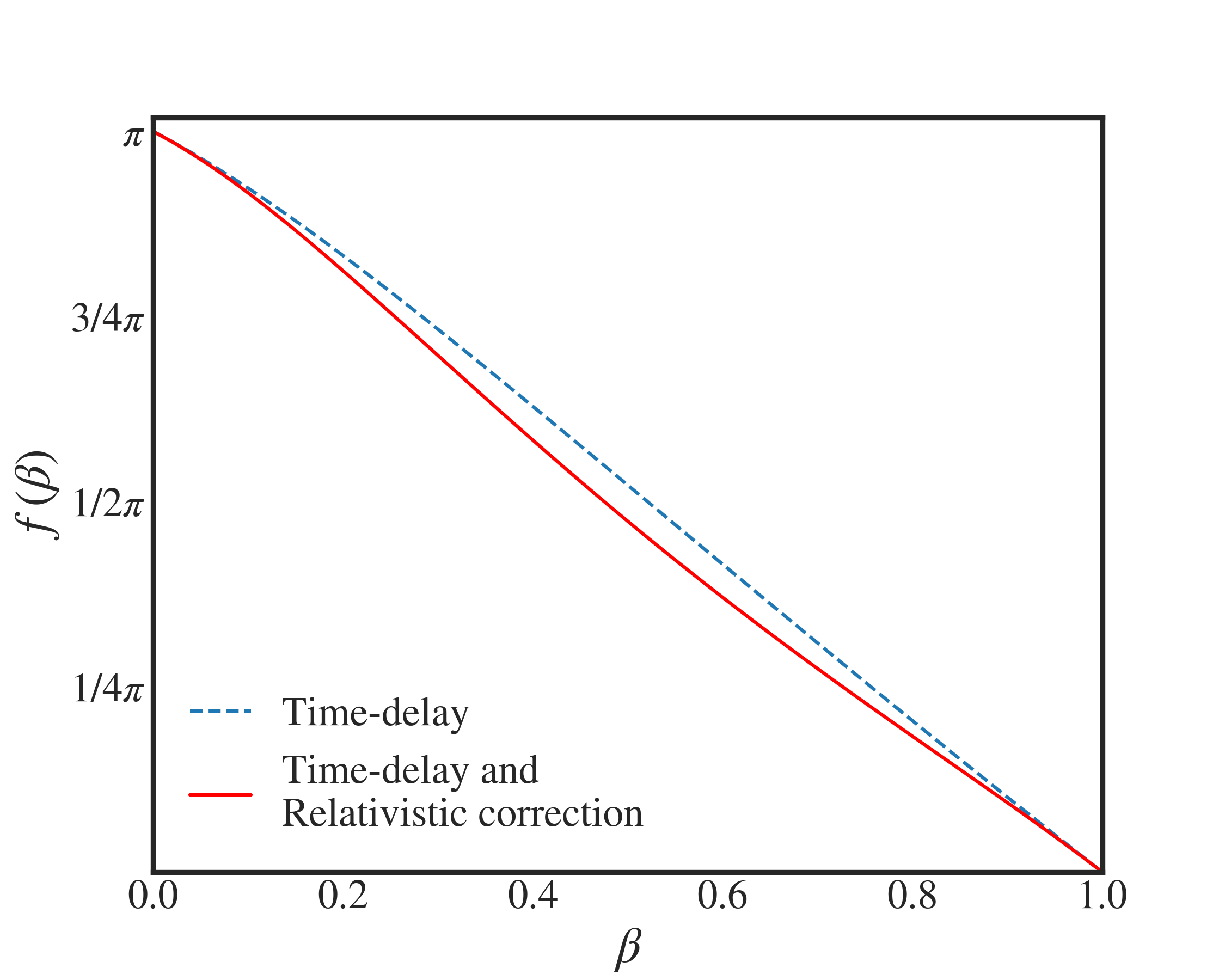}
    \caption{Time-delay and relativistic correction $f(\beta)$ as a function of expansion velocity $\beta$. For non-relativistic velocities the integral over solid angle reduces to $\pi$, but decreases at greater velocities due to the geometrical effect in \cite{Rees1967}. The blue dashed line indicates the geometrical correction only, while the red line includes the full correction of both time-delays and relativity. In the highly relativistic limit the time-delay effect implies the observed surface area approaches 0, as only the ejecta expanding towards the line of sight is visible. }  %
    \label{fig:corr}
\end{figure}%
where $f(\beta)$ represents the relativistic and time-delay correction drawn in Fig \ref{fig:corr}. The limits of integration follow from the relativistic transformation of angles (see \citealt{Sadun1991}). 

\subsection{Line relativistic correction}

The standard P~Cygni prescription codes often used in modelling lines, including the elementary supernova model, are inherently non-relativistic. It assumes an isotropic source function, an optical depth $\tau$ which is symmetric across angles, $\mu = \cos(\theta)$, and a planar resonance surface. These assumptions are valid in the co-moving frame of the ejecta, but are problematic in the observer's frame for a relativistic expansion, where aberration of angles produces an anisotropic optical depth and source function, while time dilation results in a convex surface of equal frequency. Therefore to determine the magnitude of these relativistic corrections in the mildly relativistic regime of AT2017gfo, we have also used the relativistic corrections for P~Cygni profiles derived in \cite{Hutsemekers1990}. The derived relativistic transformations are as follows. 
Firstly, the source function $S_{\rm obs}$ in the observer frame is beamed compared to the co-moving frame: $S_{\rm obs} = S_{\rm co}\,\delta(\mu)^2$ (for the relativistic transformation of the initial specific intensity $I$, see  Sect.~\ref{sec:continum}). 
Secondly, given homologous expansion, the optical depth in the observer frame, $\tau_{\rm obs}(r,\mu)$, is related to the optical depth in the co-moving frame $\tau_{\rm co}(r)$ as 
\begin{equation}
    \tau_{\rm obs}(r,\mu) = \tau_{\rm co}(r)\,\left\lvert \frac{(1-\mu \beta)^2}{(1-\beta) [ \mu (\mu - \beta)+(1-\mu^2)(1-\beta^2)]} \right\rvert
.\end{equation}Thirdly, the surface of equal frequency $\nu$ is shifted relative to the rest frequency of the line $\nu_0$ by the relativistic Doppler correction, so it must satisfy the relation 
\begin{equation}
    \frac{\nu_0}{\nu} = \Gamma(r) [1-\mu \beta(r)],
\end{equation}
which for any frequency $\nu$ can be solved for position $(r, \mu)$. \newline

\begin{figure}
    \centering
    \includegraphics[width=\linewidth,viewport=15 20 510 420,clip=]{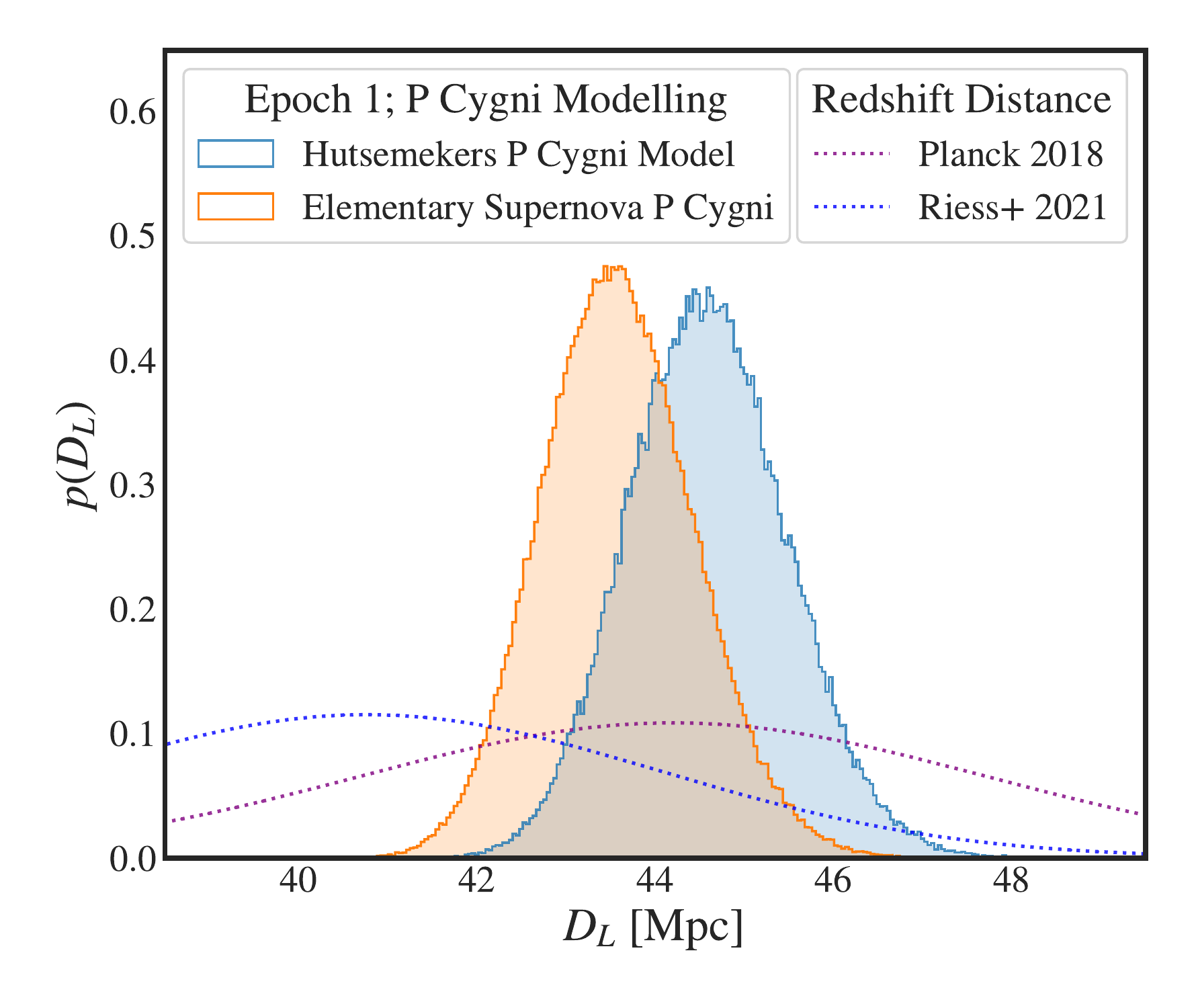}
    \caption{Effect of relativistic correction on   posterior probability distributions of luminosity distances for the first epoch spectrum. The elementary supernova model is shown in orange and the same model including the relativistic corrections of \citet{Hutsemekers1990} is shown in light blue. For the mildly relativistic regime these systematic corrections are within the uncertainties of the distance. Cosmological redshift distances are also plotted for \(H_0=67.36\pm0.54\) \citep{Planck2018} (dotted purple) and \(H_0=73.03\pm1.04\) \citep{Riess2021} (dotted blue).}
    \label{fig:variety_PCygni}
\end{figure}

These relativistic corrections are relatively minor, yielding a slight systematic increase in the inferred distance with respect to the non-relativistic P~Cygni code, but within the 90\% error region (Fig.~\ref{fig:variety_PCygni}). Given the peculiar velocity uncertainties for AT2017gfo, these relativistic corrections are minor for the current cosmological measurements, though they do affect the distance estimate by about 2\%. While we do not apply the correction here as it would be inconsistent with our line shape modelling from \cite{Sneppen2023}, for future work these effects should be included. 

\begin{table}[]
    \centering
    \caption{Characteristic contributions to the error budget for the measurement of the Hubble constant with AT2017gfo.}
    \label{table:error_budget}
    \begin{tabular}{@{}l c@{}}
     \hline\hline
        Contributor & Fractional \\
          & Uncertainty \\
    \hline
        Peculiar Velocity & 5\% \\
        Flux Calibration &  2\% \\
        Dust Extinction &   1.5\% \\
        Geometrical Constraints & 1\% \\
      \hline\hline
    \end{tabular}\label{Table:1}
\end{table}

\subsection{Summary of systematics and corrections}
We   propagated the uncertainties associated with dust extinction, flux calibration, and sphericity into our constraints on the distance (see Table \ref{Table:1}). The overall derived distance is robust to the later-emerging emission features and to the relativistic and time-delay corrections to the continuum and P~Cygni profile. The model-dependent dilution factors necessary for EPM analysis of core-collapse supernovae are expected to be and appear to be negligible in the \rprocess-dominated atmospheres of kilonovae. 
However, the methodological framework presented here, while having the virtue of simplicity, does not include a full 3D code that models the complexities of the radiative transfer self-consistently or reverberation effects. For now, we leave these higher order complexities for future simulation-based analysis. 
We note that while the blackbody nature of the continuum has so far proven challenging to reproduce in radiative transfer models from first principles, it remains an observationally motivated ansatz in this analysis.

\section{Expected constraints with future detections}\label{sec:future}

Ultimately, the analysis of the kilonova AT2017gfo cannot yield an estimate of $H_0$ decisive to the current Hubble tension due to the large peculiar velocity uncertainty of NGC\,4993. However, with  additional high-quality kilonova observations expected from the upcoming O4 and O5 runs using the Advanced LIGO, Virgo and KAGRA (HLVK) network, significantly tighter constraints are expected for a few key reasons \citep{Chen2018}. First, with larger sample size the statistical uncertainty will decrease, with the random direction of peculiar velocities averaging out. Second, due to increased GW sensitivity, future kilonova are expected mostly to lie at greater distances, where the fractional impact of peculiar velocity uncertainties are likely to be smaller. Ongoing peculiar velocity surveys will improve the accuracy of peculiar velocity models \citep{Kim2020}, likely for more distant galaxies as well. Conversely, larger distances will result in fainter optical counterparts with correspondingly lower signal-to-noise ratio (S/N) spectra, though for AT2017gfo spectral S/N is not the limiting factor in the precision of the constraints. In the following, we provide a rough estimate on the potential $H_0$ constraints attainable with this method for AT2017gfo-like kilonovae. It is probable that a significant fraction of future kilonovae will be different to AT2017gfo \citep{Gompertz2018,Rossi2020,Kawaguchi2020,Korobkin2021}, though in exactly what parameters and how they relate to their use as a $H_0$ probe is unclear. Whether the geometry of AT2017gfo is characteristic of kilonovae in general is unknown, for example, and theoretically we may expect a certain range of geometries depending on the binary parameters. Future detections of even aspherical kilonova ejecta can still be used within the EPM framework if the geometry is well constrained. Constraints on geometries can be obtained with spectropolarimetry, EPM, or line shape measurements with inclination angle constraints, complemented by numerical simulations. It is also possible that certain kilonova sub-types may be spherical, and these can be identified with such measurements of a large enough population.

\begin{figure}
    \centering
    \includegraphics[width=\linewidth,viewport=5 10 555 450,clip=]{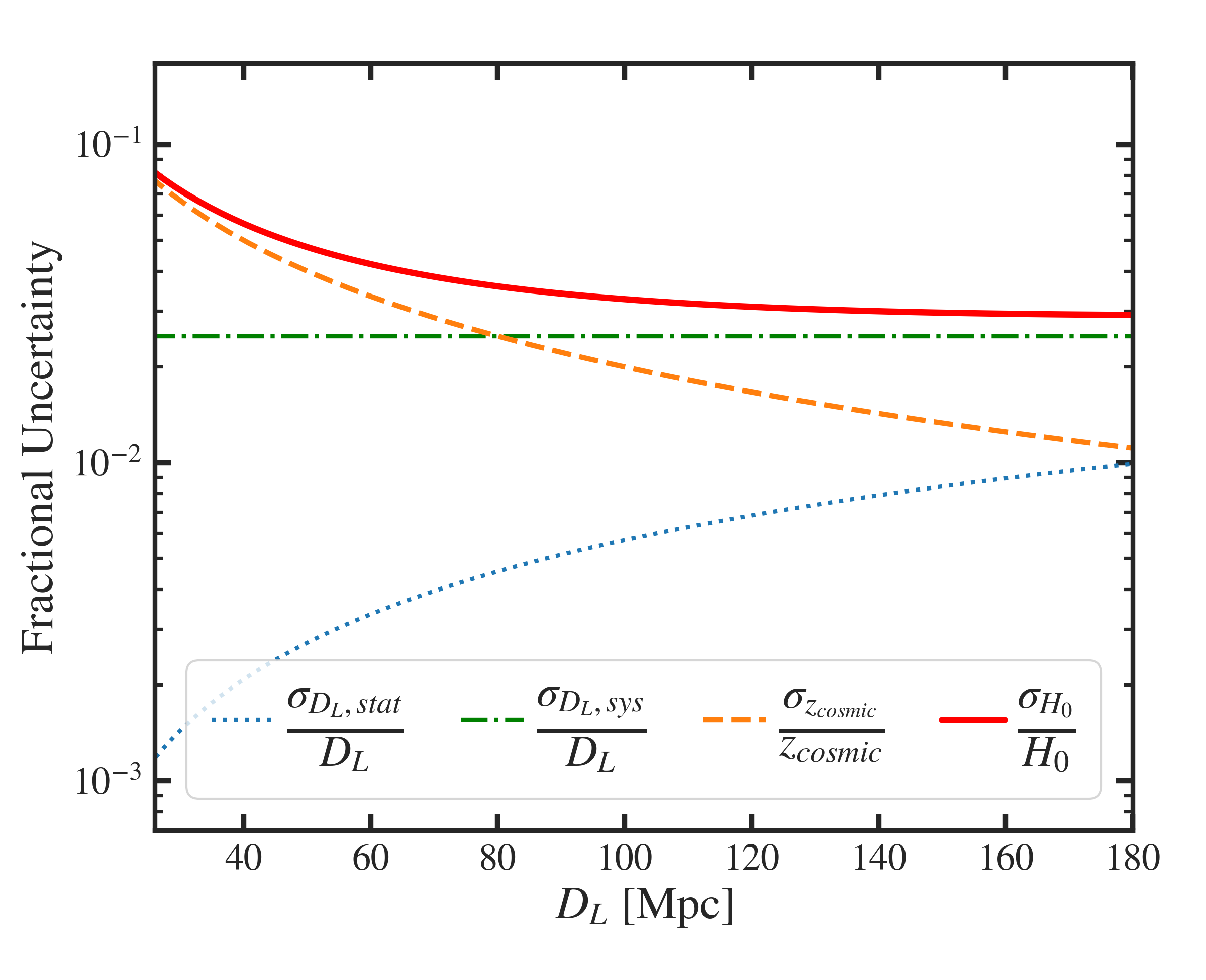}
    \caption{ Fractional uncertainty on $H_0$ (red line) for a future kilonova detection with similar intrinsic luminosity, systematic uncertainties, and peculiar velocity uncertainty as AT2017gfo. The dominant contribution for objects at $D_L<100$ Mpc is the uncertainty from peculiar velocities (dashed orange line) assuming the total peculiar velocity uncertainty is independent of the distance. The statistical uncertainty on the luminosity distance for mock spectra objects with the same intrinsic luminosity as the first X-shooter epoch increases with distance (blue dotted lines). The systematic uncertainties in distance from sphericity constraints, flux calibration, and dust extinction are considered to be independent of the source distance.  } 
    \label{fig:SNR_sim_tot}
\end{figure}

To quantify the statistical distance uncertainty the EPM framework will yield for future detections at greater distances, we generate and fit simulated spectra at variable S/N based on the AT2017gfo data. In Fig.~\ref{fig:SNR_sim_tot} we indicate the fractional uncertainty on luminosity distance obtained from a single object with an identical intrinsic luminosity as the first epoch spectrum, but placed at greater distance. Thus, the mock object at 43\,Mpc has an identical S/N valute to  the observed first epoch spectrum of AT2017gfo. Naturally, the statistical uncertainty of the fit increases monotonically, reaching a fractional uncertainty of $\approx 1\%$ for a detection at 150\,Mpc. 

Additionally, the systematic distance uncertainties associated with flux calibration, dust extinction, or the degree of asphericity contribute to the error budget. This is laid out in Table~\ref{table:error_budget} for AT2017gfo specifically. For the more general case, the relative contribution of each of these is set by the favorability of observations; for instance, the quality of coincident photometry for flux calibration, the dust extinction toward the object, or the power of line shape (or spectropolarimetry) analysis to constrain the asphericity. For AT2017gfo these yield a fractional distance uncertainty of around 2\% with the dominant systematic uncertainties being the constraint on the ejecta sphericity and the flux calibration, which are largely independent of distance. Thus, for future kilonova detections at greater distance, the fractional uncertainty due to systematics is assumed in this simple model to be unchanged (see Fig.~\ref{fig:SNR_sim_tot}).  

In contrast, to first approximation the typical peculiar velocity uncertainty is largely independent of distance, so the fractional uncertainty is assumed to decrease as $D_L^{-1}$. We note that this is a somewhat optimistic assumption as the peculiar velocity reconstruction may deteriorate slightly for distances greater than $100$\,Mpc. 
Secondly, we assume the total peculiar velocity uncertainty of NGC\,4993 is representative of future kilonova host galaxies. 
\cite{Carrick2015} predicted that the typical error when modelling the peculiar velocity field within approximately\ 250\,Mpc is around 150\,km/s. Thus, the NGC\,4993 peculiar velocity error, 130\,km/s, is quite typical, in no way particularly large or small. 
Subject to these two assumptions, the peculiar velocities remain the dominant uncertainty up to $D_L \approx 80$\,Mpc, as illustrated in Fig.~\ref{fig:SNR_sim_tot}. An individual kilonova observation at $D_L \approx 150$\,Mpc with both GW signatures and high S/N spectral follow-up could alone yield an estimate of $H_0$ to $\sim3$\% precision.

The largest unknown in forecasting future constraint capabilities for HLVK runs O4 and O5 is estimating the number of well-localised binary neutron star (BNS) mergers (or at least, those similar to AT2017gfo). We do not comment on this issue here. 
Instead in Table~\ref{table:future} we show the expected constraint on $H_0$ attainable with one, five, and ten kilonovae  with similar intrinsic luminosity and peculiar velocity to AT2017gfo. These detections are assumed to be uniformly distributed within the observable volume up to 150\,Mpc. The typical BNS merger for this volume is located at $D_L = 119_{-36}^{+22}$\,Mpc. It is worth noting that within these distances, the cosmic variance for any local measurement of $H_0$ is at the one percent level \citep{Wojtak2014}, which sets a lower bound on the cosmological precision attainable from averaging over local objects. Nevertheless, these results suggest that applying EPM to kilonovae may yield percent-precision $H_0$ constraints given the current detection sensitivities and a handful of detections. For the nearest kilonova observations, at smaller or similar distances to NGC\,4993, this framework can be complemented by, and even incorporated within, the cosmological distance ladder, for example   by making Cepheid observations within the host galaxies and comparing them with the inferred distance.

\begin{table}
\caption{Future number of kilonovae detected with GW and optical counterparts vs the expected fractional constraint attainable on $H_0$. All mergers are assumed to be uniformly distributed within an observable volume with a distance of 150\,Mpc, with identical intrinsic luminosities and peculiar velocity uncertainties as AT2017gfo.}\label{table:future}
\renewcommand{\arraystretch}{1.5} 
\centering
\begin{tabular*}{0.4\textwidth}{c @{\extracolsep{\fill}} ccc}
\\ \hline \hline 
BNS mergers     & 1 & 5 & 10 \\ \hline 
$\frac{\sigma_{H_0}}{H_0}$  &  $3 \%$  & $1.4 \%$ &  $1.0 \%$ \\ 
\hline \hline
\end{tabular*}
\end{table}

\section{Conclusions}
We have shown in this paper how spectral analysis of well-observed kilonovae can be used to measure the expansion rate of the universe to a precision of a few percent, with low systematic uncertainty. Using AT2017gfo as an example, we have examined the dominant statistical and systematic uncertainties and derived a preliminary value of the Hubble constant accurate to about 5\%, dominated by the host galaxy peculiar velocity. While we believe that the important systematic effects have been considered, we reiterate that the value of $H_0$ we present here is preliminary and may be subject to significant change once a full 3D radiative transfer simulation can accurately capture effects such as the light travel time reverberation and the true nature of the kilonova atmosphere, which are not accounted for in our modelling. We also show that a sample of a handful of kilonovae with  luminosity and characteristics similar to  AT2017gfo, such as may be detected in the upcoming HLVK runs O4 and O5, and well observed spectroscopically, could be made to yield a value of $H_0$ accurate to about 1\%.  

\section{Acknowledgements}
We thank Jonathan Selsing, Stuart Sim and Ehud Nakar for useful discussions and feedback. We also wish to express gratitude to Brandon Hensley, Bruce Draine, Douglas Finkbeiner, Anja Andersen, and Adam Riess for providing comment to improve the discussion on uncertainties from dust extinction.
The Cosmic Dawn Center is funded by the Danish National Research Foundation under grant number 140.  
AB and OJ acknowledge support by the European Research Council (ERC) under the European Union's Horizon 2020 research and innovation programme under grant agreement No. 759253, by the Deutsche Forschungsgemeinschaft - Project-ID 279384907 - SFB 1245, and by the State of Hesse within the Cluster Project ELEMENTS. RW was supported by a grant from VILLUM FONDEN (project number 16599)


\bibliographystyle{mnras}
\bibliography{refs} 

\end{document}